# A Mass Measurement for the Missing Baryons in the Warm-Hot Intergalactic Medium via the X-ray Forest

Fabrizio Nicastro*, Smita Mathur†, Martin Elvis*, Jeremy Drake*, Taotao Fang‡, Antonella Fruscione*, Yair Krongold*††, Herman Marshall‡‡, Rik Williams†, Andreas Zezas*

*Harvard-Smithsonian Center for Astrophysics 60 Garden St, Cambridge MA 02138

†Astronomy Department, The Ohio State University, 43210, Columbus, OH, USA

‡ University of California-Berkeley, Berkeley, CA, USA (Chandra Fellow)

††Istituto de Astronomia, UNAM, Mexico City, Mexico

‡‡Massachusetts Institute of Technology, Cambridge, USA

**Recent cosmological measurements indicate that baryons comprise about four percent of the total mass-energy density of the Universe[1,2], which is in accord with the predictions arising from studies of the production of the lightest elements[3]. It also is in agreement with the actual number of baryons detected at early times (redshifts $z>2$)[4,5]. However, close to our own epoch ($z<2$), the number of baryons actually detected add up to just over half (~55 percent) of the number seen at $z>2$ ([6-11]), meaning that about ~45 percent are 'missing'. Here we report a determination of the mass-density of a previously undetected population of baryons, in the warm-hot phase of the intergalactic medium. We show that this mass density is consistent, within the uncertainties, with the mass density of the missing baryons.**

Hydrodynamical simulations for the formation of structures in the Universe offer a possible solution to the puzzle of the ``missing baryons" (e.g. [12,13,14]). They predict that in the present epoch (z ≲1-2), ∼ 30-40 % of the normal baryonic matter in the Universe lies in a tenuous Warm-Hot Intergalactic Medium (WHIM) at overdensities $\delta = n_b / <n_b> \approx$ 5-100 ($<n_b> = 2 \times 10^{-7} (1 + z)^3 (\Omega_b h^2/0.02)$ cm$^{-3}$ is the average density in the Universe), a factor $10^6$ below the density of the interstellar medium in our Galaxy. The WHIM was shock-heated to temperatures of $10^5$-$10^7$ K during the continuous process of structure formation[12,13]. At these temperatures C, N, O and Ne (the most abundant metals in gas with solar or solar-like composition), are highly ionized, being mainly distributed between their He-like and H-like species. The strongest bound-bound transitions from these ions fall in the soft X-ray band (e.g. $\lambda(OVII_{K\alpha})$ = 21.602 Å, and $\lambda(OVIII_{K\alpha})$ = 18.97 Å), and so a 'forest' of weak (i.e. N < $10^{16}$ cm$^{-2}$, or W < 18-30 mÅ depending on the particular ion) absorption lines is expected to be imprinted in the X-ray spectra of all background sources at redshifts between zero and ∼ 1-2 by these WHIM filaments (e.g. [14,15]), as direct analogs of the 'Lyman-α forest' due to cool neutral hydrogen detected in the optical spectra of quasars. The Lyman-α forest accounts for virtually all the baryons at $z > 2$[4,5], but only for about 30% of them at lower $z$[9] (Table 1). Due to our special embedded location, metal absorption lines from warm-hot gas at z=0 (either Local Group WHIM[16] or an extended Galactic halo[17,18]) has been detected in X-rays[19,20,21,22] and in the UV[16,17]. However, even if all this matter were to be identified with Local Group WHIM, our preferential location would not allow us to infer the WHIM contribution to the cosmological baryon density in the local Universe. Measuring $\Omega_b$(WHIM) requires observing along a random, unprivileged line of sight. Due to the intrinsic steepness of the number of absorbers as a function of ion column density for the WHIM metal absorption lines, they are rare at relatively large columns (e.g. [14]). Only a single OVII Kα system with $N_{OVII} \geq 10^{16}$ cm$^{-2}$ is expected along a random line of sight, up to $z = 0.3$, with a probability of ~ 60 %, while 8 systems with $N_{OVII} \geq 10^{15}$ cm$^{-2}$ are expected. So detecting weaker lines is expected to be more productive than probing higher redshifts.

In practice a very large number of counts per resolution element is needed to detect these lines in current *Chandra* Low Energy Transmission Grating (LETG) or XMM-*Newton* Reflection Grating Spectrometer (RGS) spectra, both of which have a resolution of 50 mÅ (R ~ 400 at ~20 Å). To probe the more common WHIM columns of $N_{OVII}$ ~ $10^{15}$ cm$^{-2}$, which have equivalent width of only ~ 3 mÅ, about 2500 counts per resolution element are needed for a 3σ detection of a non-saturated line. The brightest Seyfert galaxies and Blazars have typical soft X-ray (0.5-2 keV) fluxes of the order of 4 x $10^{-11}$ erg s$^{-1}$ cm$^{-2}$ (2 mCrab), which would take a 2.5 Ms *Chandra*-LETG exposure to reach this S/N. Furthermore all these objects are nearby ($z \lesssim 0.05$) and so sample path-lengths of only a few hundred Mpc. Indeed no secure X-ray detection of higher redshift intervening WHIM absorption systems has been claimed so far[23]. One solution is to observe background sources in unusually bright states.

On 2002, October, 26-27 and 2003, July 1-2 we triggered, under our Chandra-AO4 program, two 100 ks LETG observations of the Blazar Mkn 421. These two observations caught the source at historical maxima of 60 and 40 mCrab in the 0.5-2 keV band, allowing us to collect the best S/N of any high resolution spectrum taken with *Chandra* to date, including Galactic X-ray binaries. We detected 24 metal absorption lines in the LETG spectrum of Mkn 421 (Figure 1), with conservative significances > 2σ (for additional details, please see the Supplementary Methods – SM – available online). We identify 9 of these 24 absorption lines as belonging to 2 different intervening absorption systems, at $z = (0.011 \pm 0.001)$ and $z = (0.027 \pm 0.001)$ (Table 2). The remaining lines are identified as belonging to either the interstellar medium (ISM) of our Galaxy or the Local group WHIM Filament, and will not be discussed further here.

An intervening HI Lyα system at $cz = (3046 \pm 2)$ km s$^{-1}$ ($z=0.010160 \pm 0.000007$) was discovered in the Hubble Space Telescope (HST) GHRS spectrum of Mkn 421, and reported by Shull et al[24]. The redshift of the HI Lyα line is consistent with that of the $z=0.011$ X-ray absorption system within the large systematic X-ray uncertainties ($\Delta z = 0.001$), but its FWHM implies a 3σ upper limit on the temperature of the HI absorber of $T^{HI}_{max} < 1.2 \times 10^5$ K, inconsistent with the range of temperatures estimated for the X-ray absorber (see caption of Figure 2). If the systems are related, a multiphase WHIM is required. No HI Lyα lines are visible at $z=0.011$ and $z=0.027$.

We also searched the Far Ultraviolet Spectrometer Explorer (FUSE) spectrum of Mkn 421 for OVI absorption at or close to the redshifts of our two X-ray absorbers. No OVI lines are visible at $z=0.010160$, $z=0.011$, or $z=0.027$.

The conservative significances of the X-ray absorbers at $z=0.011$ and $z=0.027$ are 3.5σ and 4.9σ respectively, while their direct integrated significances are 5.8σ and 8.9σ (Table 2).

For both the $z > 0$ X-ray systems we compared the measurements (X-rays) or 3σ upper limits (UV, FUV and X-rays) of the equivalent widths and equivalent width ratios with ion relative abundances and ratios from hybrid collisional-ionization plus photoionization models (see SM for details), to derive the physical properties of the gas (e.g. [19]).

For both systems we find solutions that define physical, geometrical and astrophysical parameters consistent with theoretical predictions for the WHIM (see caption of Figure 2 for details). We therefore identify these two intervening absorbers with two tenuous WHIM filaments absorbing the UV and X-ray radiation of Mkn 421 along our line of sight (see the online Supplementary Discussion - SD - for a discussion on the arguments used to rule out the identification of the absorbers with reprocessed material outflowing from the blazar environment or its host galaxy).

Previous works have attempted to estimate the baryon content of the OVI and OVIII WHIM based on either X-ray absorption line upper limits[26] or single-line detections[27] (none of which were confirmed by subsequent observations[19,22]). These estimate all suffer the large uncertainties due to the poor ionization corrections used and the absence of a metallicity estimate. Based on the two WHIM detections presented here we can, for the first time, estimate the number density of OVII WHIM filaments with $N_{OVII}^{Thresh} \geq 7 \times 10^{14}$ cm$^{-2}$ (the weakest OVII column that we detect, in the z=0.027 filament) in the local Universe. We find $d\mathcal{N}/dz = 67_{-43}^{+88}$, which is roughly 4 times the number density of the ~ 17.5 times weaker OVI absorbers[28], and is consistent within the 1σ errors[29] with hydrodynamical simulation predictions, $d\mathcal{N}/dz$(expected) = 30.1 (Figure 2)[13,14].

More importantly, the accurate ionization correction and metallicity estimates enabled by OVI-OVIII, NVI-VII and HI ratios, allow us to derive a first estimate of the mean cosmological baryon density in X-ray filaments. Following Savage et al.[28] (and approximating their formula 1 to the low-redshift regime), we estimate the baryon mass density of the WHIM, in units of critical density $\rho_c$ as: $\Omega_b = \left(\dfrac{1}{\rho_c}\right)\left(\dfrac{\mu m_p \sum_i N_H^i}{d_{Mkn421}}\right)$, where μ is the average molecular weight (taken to be 1.3), $N_H^i$ is the equivalent H column density of the WHIM filament $i$ and $d_{Mkn421} = (zc/H_0) = 128$ $h_{70}^{-1}$ Mpc is the distance to Mkn 421. Adopting a central temperature of logT(K) = 6.1 for both systems, gives $\Omega_b^{WHIM} = (2.7_{-1.9}^{+3.8}\%) \times 10^{-[O/H]_{-1}}$ (here we included both the measured uncertainties on $N_H$ and the small number statistical errors for a Poissonian distribution[29], but did not attempt to estimate the size of the uncertainty due to cosmic variance, since we have only observed a single line of sight). This is consistent with the expected $\Omega_b^{WHIM}(predicted) = 1.6 - 1.8\%$ [13,14], and with the total amount of 'missing' baryons $\Omega_b(missing) = (2.1^{+0.5}_{-0.4})\%$ (Table 1).

**Supplementary Information accompanies the paper on www.nature.com/nature**

**Correspondence and requests for materials should be addressed to F.N. (fnicastro@cfa.harvard.edu)**

Acknowledgements

The authors thank two anonymous referees for useful comments and suggestions that helped improve the paper substantially. This work has been partly supported by NASA-Chandra and NASA-LTSA grants.

**Competing interests statement**
**The authors declare that they have no competing financial interests**

**Table 1**

| Inferred From | $\Omega_b$(%) for $h_{70}=1$ | References |
|---|---|---|
| BBN +D/H | $(4.4 \pm 0.4)$ | [3] |
| CMB Anisotropy | $(4.6 \pm 0.2)$ | [1,2] |
| **Observed at *z* > 2 in** | | |
| Lyman-$\alpha$ Forest | > 3.5 | [4,5] |
| **Observed at *z* < 2 in** | | |
| Stars | $(0.26 \pm 0.08)$ | [6] |
| HI + HeI + H$_2$ | $(0.080 \pm 0.016)$ | [6] |
| X-ray gas in Clusters | $(0.21 \pm 0.06)$ | [6] |
| Lyman-$\alpha$ Forest | $(1.34 \pm 0.23)$ | [9] |
| Warm + Warm-Hot OVI | $(0.6^{+0.4}_{-0.3})$ | [10,11] |
| Total (at *z* < 2) | | $(2.5^{+0.5}_{-0.4})$ |
| Missing Baryons (at *z* < 2) | | $(2.1^{+0.5}_{-0.4})$ |

**Census of Baryons in the high and low redshift Universe.** Rows 2-3 list the cosmological baryon density (Col. 2) inferred by Big Bang Nucleosynthesis compared with observations of light element ratios, and observations of CMB anisotropies, respectively (references are given in Col. 3). Rows 5 lists the cosmological baryon density found in the Lyman-$\alpha$ Forest at z > 2 (Col. 2) and reference (Col. 3). Virtually all baryons are found in this cool component at high redshift. Rows 7-11 list the cosmological baryon density (Col. 2) observed at *z* < 2 in virialized (stars, neutral H and He and molecular H and X-ray emitting hot gas in clusters of galaxies and big groups -M > 4 $10^{13}$ M$_\odot$, respectively) and non yet virialized (cool gas in the local Lyman-$\alpha$ forest and warm-hot gas in the OVI WHIM) structures. References are in Col. 3. Finally the last two rows give the total observed cosmological baryon density at z < 2 and its difference from the cosmological baryon density inferred by observations of CMB anisotropies (row 3, col. 2), i.e. the so called 'missing' baryons.

**Table 2**

| Redshift | >2σ Ion Detections | Probability of Exceeding F | Conservative Significance | Sum of Lines Significance |
|---|---|---|---|---|
| 0.011 ± 0.001 | OVII, NVII | $5.9 \times 10^{-4}$ | 3.5σ | 5.8σ |
| 0.027 ± 0.001 | OVII, NVII, NVI | $1.3 \times 10^{-6}$ | 4.9σ | 8.9σ |

**The two WHIM Filament along the line of sight to Mkn 421.** Column 1 lists the redshift of the two X-ray absorbers and its error, which include a systematic 20 mÅ uncertainty in the LETG wavelength calibration. Column 2 lists the ions detected at a significance >2σ. CVI is detected in the z=0.027 system, at a significance of 1.8σ. OVIII and NeIX are detected in the z=0.011 and z=0.027 systems respectively, at significance larger than 3σ, but both detections are considered here as 3σ upper limits due to the proximity of these lines with known instrumental artifacts. The total conservative significance of the two systems (Col. 4) was estimated by comparing the best fit $\chi^2$ of simpler models not including the metal absorption lines from these systems, with that of more complex models that include them. The F-tests were run over the following two portions of the spectrum, containing a similar number (~100) of resolution elements: Δλ(z=0.011) = [21,26] Å and Δλ(z=0.027) = [21.5,23]∪[24.5,26] ∪ [27.5,30] ∪ [34,35]. $OVII_{k\alpha}$ and $NVII_{k\alpha}$ lines were included in the complex model for the z=0.011 system, while $OVII_{k\alpha}$, $NVII_{k\alpha}$, $NVI_{k\alpha}$ and $CVI_{k\alpha}$ were used for the z=0.027 system. The width of the lines was frozen to the instrument resolution, while their relative positions and normalizations were fixed to (a) their expected relative position based on the proposed identifications, and (b) their measured relative normalization. Only the redshift of the system and the overall normalization were left free to vary in the fits. These F-tests yielded the probabilities listed in Column 3, and so the significances listed in Column 4. Finally, Column 5 lists the sum of the significances of the lines listed in Column 2 (each computed as ratio between measured EW its 3σ error: see SM).

**Figure 1**

**The WHIM absorption in the Chandra LETG spectrum of Mkn 421.** Six portions of the LETG spectrum of MKn 421 along with its best fitting continuum plus narrow absorption model (solid line), centered around the rest-wavelengths of the (a) NeIX-X K$\alpha$, (b) OVIII K$\alpha$, (c) OVII K$\alpha$, (d) NVII K$\alpha$, (e) NVI K$\alpha$ and (f) CVI K$\alpha$ transitions. This spectrum contains a total of ~ 5000 counts per resolution element in the continuum at 21 Å, enough to detect OVII columns of $N_{OVII} \gtrsim 8 \times 10^{14}$ cm$^{-2}$ at a significance level $\geq$ 3$\sigma$. For each labeled ion in the six panels, the three vertical lines from left to right are the rest frame wavelengths (thin line) and the expected wavelengths at z=0.011 (*cz*=3300 km s$^{-1}$, medium thickness line) and z=0.027 (*cz*=8090 km s$^{-1}$, thick line).

Mkn 421 was also observed with the HST GHRS on 1995 February 1. We retrieved this GHRS spectrum of Mkn 421 from the public HST archive, and re-analyzed the data. The 3$\sigma$ HI column upper limit of putative HI Ly$\alpha$ at the average redshifts of the two X-ray systems are $N_{HI} < 4.7 \times 10^{12}$ cm$^{-2}$ and $N_{HI} < 8.5 \times 10^{12}$ cm$^{-2}$ (assuming a temperature of logT(K) = 6.1 for both systems).

Following our TOO request Mkn 421 was observed by the Far Ultraviolet Spectrometer Explorer (FUSE) on 2003, January 19-21, with a total exposure time of 62 ks. From this spectrum we derived 3$\sigma$ OVI column upper limits of putative OVI$_{2s \rightarrow 2p}$ at the redshift of the HI Ly$\alpha$ absorber and at the average redshifts of the two X-ray systems, of $N_{OVI}$(z=0.01) < 1.4 × 10$^{13}$ cm$^{-2}$, $N_{OVI}$ (z=0.011) < 1.6 × 10$^{13}$ cm$^{-2}$ and $N_{OVI}$ (z=0.027) < 1.4 × 10$^{13}$ cm$^{-2}$ respectively.

**Figure 2**

**The 'Missing' baryons have been found in the WHIM.** Predicted curves $d\mathcal{N}_{OVII}/dz(\geq N_{OVII}^{Thresh})$ versus $N_{OVII}^{Thresh}$ [14]. The data point $[d\mathcal{N}_{OVII}/dz(N_{OVII}\geq 7x10^{14})]_{observed} = 67_{-43}^{+88}$ is also shown with $\pm 1\sigma$ errors, and is consistent with the corresponding theoretical expectation value $[d\mathcal{N}_{OVII}/dz(N_{OVII}\geq 7x10^{14})]_{expected} = 30.1$. All physical, geometrical and astrophysical parameters of the two observed intervening systems are also fully consistent with WHIM expectations. For the system at z=0.011 we find common $N_H$ solutions in a relatively broad range of temperatures: logT(K) = (5.8-6.4), and constrain the oxygen metallicity to [O/H] > -1.47 (note that, to find these solutions, we use the $3\sigma$ upper limit on the column of the putative HI Ly$\alpha$ at z=0.011, and not the measured column of the HI Ly$\alpha$ at z=0.010160, since the temperature implied by the width of this line is inconsistent with the range of temperatures that define the X-ray solution). For the system at z=0.027 we find that solutions exist only for a relatively narrow $[\Delta T \approx 3x10^5 K]$ range of temperatures around T = 1.4 x $10^6$ K, and only for rather high N/O ratios compared to Solar: [N/O] $\geq$ 0.65, and again moderately high metallicity [O/H] $\geq$ -1.32. The equivalent H column densities of the two intervening systems are $N_H(z=0.011) = [(1.7\pm 0.7)T_6 - (0.1\pm 0.6)]\times 10^{19}10^{-[O/H]_{-1}}$ cm$^{-2}$ and $N_H(z=0.027) = [(1.67\pm 0.76)T_6 + (0.71\mp 1.15)]\times 10^{19}10^{-[O/H]_{-1}}10^{-[N/O]_{0.65}}$. These, in turn, imply thicknesses of the absorbers along the line of sight of $D(z=0.011) = [(0.55\pm 0.23)T_6 - (0.03\pm 0.19)] (n_{-5b})^{-1} 10^{-[O/H]_{-1}}$ Mpc and $D(z=0.027) = [(0.53\pm 0.24)T_6 + (0.23\mp 0.37)] (n_{-5b})^{-1} 10^{-[O/H]_{-1}}10^{-[N/O]_{0.65}}$ Mpc ($n_{-5b}$ is the baryon volume density in units of $10^{-5}$ cm$^{-3}$).

**Figure 1**

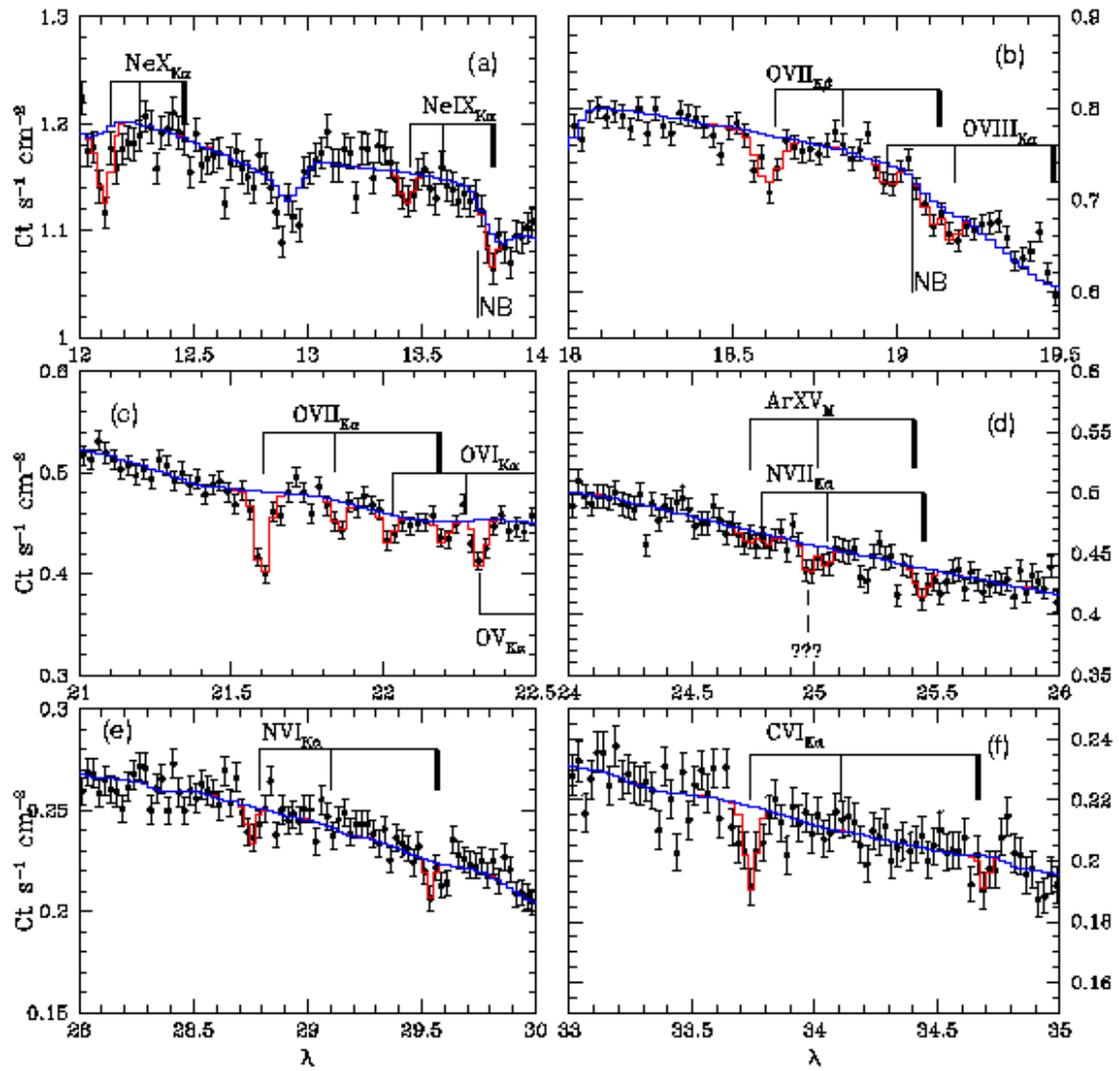

**Figure 2**

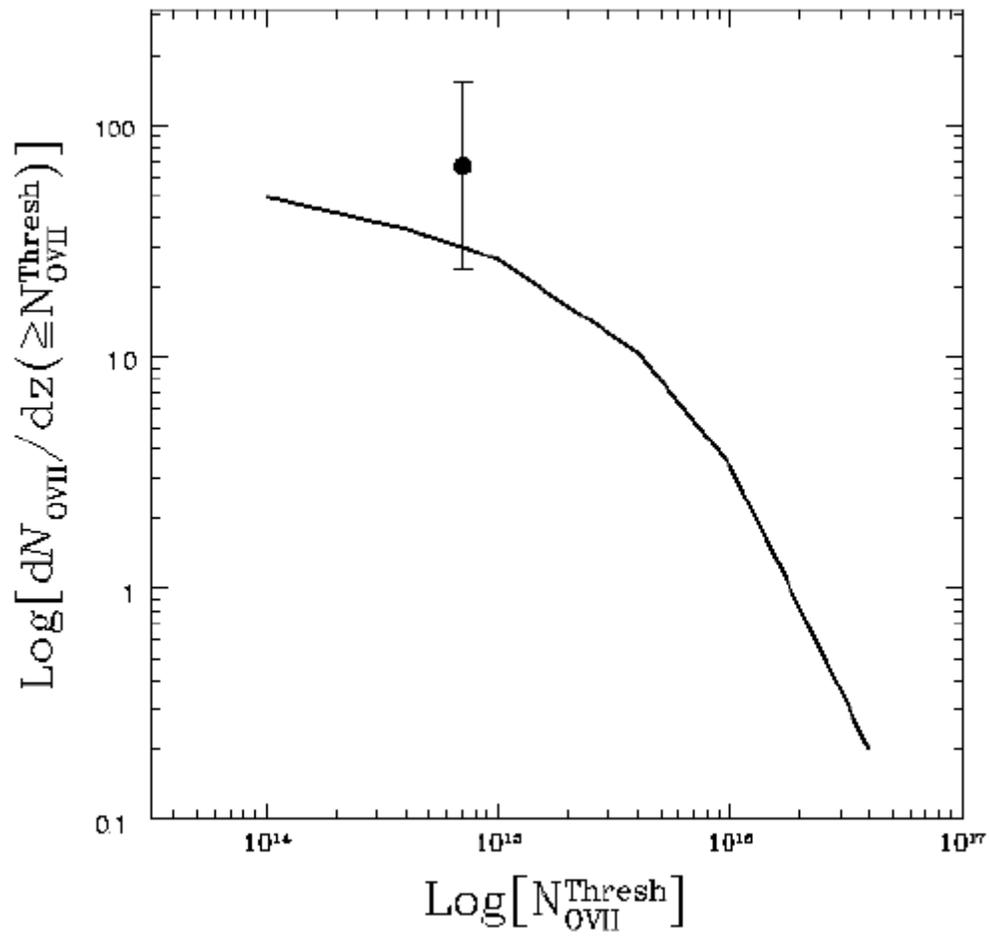

# Appendix A: Supplementary Methods

## Observations

Mkn 421 was observed three times with the Chandra LETG, on 2000, May 29 (HRC-LETG, 19.7 ks), 2002, October 26 (ACIS-LETG, Figure 1: 91.5 ks) and 2003, July 3 (HRC-LETG, 99.2 ks). These three observations were all performed as Target of Opportunity Observations (TOOs), following intense activity of the target as recorded by the Rossi-XTE All Sky Monitor (ASM). The source fluxes during these three observations, in the soft X-ray 0.5-2 keV band were 13.5, 60 and 40 mCrab respectively (1 mCrab = 2 x $10^{-11}$ erg $s^{-1}$ $cm^{-2}$).

## Data Reduction and Analysis

The data reduction and analysis was performed with the version 3.0.2 of the 'Chandra Interactive Analysis of Observations' (CIAO) software.
To increase the Signal to Noise (S/N), we combined the first order ACIS-LETG spectrum (and responses) of Mkn 421 with the two non-order-sorted HRC-LETG spectra (and responses) of the target in the overlapping 10-60 Å wavelength range. The final co-added spectrum contains more than 7 million counts, and has ~ 5300 counts per resolution element in the continuum at 21.6 Å. This S/N was enough to detect OVII columns of $N_{OVII} \gtrsim 8 \times 10^{14}$ $cm^{-2}$ at a significance level ≥ 3σ.
ACIS-LETG negative and positive $1^{st}$ order effective areas (ARFs) were built using the *fullgarf* script, with the ACIS-Quantum-Efficiency (QE) files made available by the ACIS calibration group (CALDB 2.2.6) to correct for the ACIS contamination problem[1] (e.g. [2]). Negative and positive ACIS-LETG $1^{st}$ order ARFs were then checked for consistency and combined using the Ciao tool *add_grating_orders*, and finally convolved with the standard $1^{st}$ order ACIS-LETG Redistribution Matrix (RMF, CALDB 2.2.6). For the HRC-LETG responses, instead, we used the standard convolution products of RMFs and ARFs files, for orders 1 through 6 ([3]). Finally, the first order ACIS-LETG and order 1 through 6 HRC-LETG responses were combined together (weighting by count rate) to form a ACIS-LETG+HRC-LETG response matrix.

We used the fitting package *Sherpa*[4], in CIAO, to model the broad band 10-60 Å average continuum of Mkn 421 during the three LETG observations. To isolate and correct for the ≲ 20 % amplitude instrumental artifacts, which are due to residual calibration uncertainties (particularly at the wavelengths of the OI and CI K edges) and to bad columns in the ACIS detector corresponding to CCD node boundaries or chip gaps, we added ten broad (FWHM = 0.5-7 Å) Gaussians (6 in emission and 4 in absorption) to our power law plus Galactic absorption (with varying ISM metallicities) best fit continuum model. An ApJ paper with full details is currently in preparation (Nicastro et al., 2005). Our final best fitting continuum model yielded a $\chi^2_r(dof) = 1.81(1598)$, with relatively flat broad band residuals and several narrow negative spikes. These spikes are all due to the narrow resonant absorption lines that are the subject of this paper (28 bins with ≥ 3σ negative residuals, with 16 of these being ≥ 4σ). Including 24 absorption

Gaussians to model these narrow features in the spectrum reduced the $\chi_r^2$ to $\chi_r^2(dof) = 1.25(1529)$, giving an F value of 14.2, when compared with the best fitting continuum model with no lines, corresponding to a probability of exceeding F, P(>F) << 0.001.

## Line Significance

The significance of the lines depends critically on the degree of accuracy with which the local continuum is known. To assess this dependence for the weakest lines in our spectrum we repeated locally the $\chi^2$ test on our best fitting continuum for all the 'interesting' regions plotted in Figure 2 of the main paper, after eliminating narrow (50-100 mÅ) intervals of data centered on the z=0 lines. We let the normalization of our best fitting continuum vary locally, and refitted the data. In all cases we obtained slightly lower relative normalizations (with values ranging from 0.95 to 0.98) and unacceptably high reduced $\chi^2$ values for the given number of d.o.f. As an example, the 21.4-22.5 Å interval gave a continuum relative normalization of 98.9% and $\chi_r^2(dof) = 1.4(23)$, corresponding to a probability of exceeding $\chi^2$ of $P_{cont}(>\chi^2) = 9\%$. Including Gaussians for the two lines at 21.85 Å and 22.20 Å, and refitting the data leaving again the relative continuum normalization and the 6 parameters of the Gaussians free to vary, gives a relative continuum normalization of 1.0 and $\chi_r^2(dof) = 1.0(17)$, corresponding to $P_{cont+lines}(>\chi^2) = 45\%$. An F test between the two models gives a probability of 10% of exceeding F. All other cases gave lower $P_{cont}(>\chi^2)$ and $P(>F)$.

## Comparison of Data with Models

We used CLOUDY[5] to build grids of mixed collisional ionization and photoionization models for temperatures in the range 5 < logT < 7 and for a baryon volume density of $n_b = 10^{-5}$ cm$^{-3}$. The photoionizing field includes the secondary photoionization contribution from both the blazar during outburst phases and the metagalactic UV and X-ray background at the redshifts of the two systems, parameterized as in [6].

For both the z > 0 X-ray systems we compared the measurements (X-rays) or 3σ upper limits (UV, FUV and X-rays) of the equivalent width ratios, with ion relative abundance ratios from our grid of hybrid collisional-ionization plus photoionization models, to constraint maximal intervals of temperatures defining common solutions (e.g. [6]). The existence of a common range of temperatures for all ion ratios still does not guarantee that a range of self-consistent equivalent H column density solutions exist. To search for these solutions we compared the absolute value of the measured equivalent widths for each observed ionic transition, with the predicted values within the maximum allowed range of temperatures. This defined two ranges of self-consistent $N_H$ versus T solutions for the two systems, and constrained their metallicity. We note that although our estimates of the maximal intervals of temperatures (based on the observed ion equivalent width ratios) depend slightly on the significance level used for the upper limits of the HI and OVI equivalent widths, maximal intervals of temperatures can still be found for both systems even using only 1σ upper limits for these two ions, and these are still of the order of the final ranges of temperatures that define the self-consistent ($N_H$,T) solutions. In particular, for the *z*=0.011 system, the lower boundary of the maximal temperature range (defined by the $N_{OVI}/N_{OVII}$ ratio) would increase from logT =5.72 to logT=5.82 going from the 3σ to the 1σ significance level for $N_{OVI}$. This is of the order of the lower boundary found for the self-consistent ($N_H$,T) solutions for this system (logT=[5.8,6.4]: see caption of Figure 2 in the main article). Analogously, for the z=0.027 system, the upper boundary of the maximal temperature range (defined by the $N_{HI}/N_{NVII}$ ratio) would decrease from logT=6.5 to logT=6.15, going from the 3σ to the 1σ significance level for $N_{HI}$. This is again of the order of the upper boundary found for the self-consistent ($N_H$,T) solutions for this system (logT=[6.1,6.2]: see caption of Figure 2 in the main article). The significance level at which we consider the upper limits of $N_{HI}$ and $N_{OVI}$ does not affect our conclusions.

## References to Appendix A

# Appendix B: Supplementary Discussion

## Line Identification: ruling out Blazar Outflows

The 24 lines in the spectrum are identified with three systems: at z=0 (including both neutral and ionized ISM as well as highly ionized Local-Group WHIM[1,2,3,4,5] or extended Galactic halo[6,7]), z=0.011 and z=0.027 (two intervening WHIM filaments). For all 9 z>0 identifications we considered the possibility of misidentification with either high oscillator strength outer shell transitions from less abundant elements (i.e. Ar and Ca)[8,9] or poorly known z=0 inner shell transitions from abundant metals (C, N, Ne and O)[9,10,11]. We concluded that at most one of the 9 detected lines, at $\lambda = 13.80$ Å, can be tentatively identified with a z=0 line from inner-shell NeVIII. All other lines must belong to redshifted systems (see Nicastro et al., 2004, in preparation).

We investigated the possibility that these lines were imprinted on the spectrum by highly ionized material outflowing from either the blazar environment or the blazar's host galaxy with velocities of ~ 5600 km s$^{-1}$ and ~ 900 km s$^{-1}$, rather than by intervening WHIM filaments at z=0.011 and z=0.027. Even the lower of these two velocities is extreme for ISM clouds, when compared with the typical range of velocities observed in High Velocity Clouds (either cold, e.g. [12] or warm-hot, e.g. [6]) in our Galaxy. However, photoionized outflows with velocities ranging between few hundreds and 1000-2000 km s$^{-1}$ are common in low redshift Seyfert 1s (e.g. [13]). We note, however, that all these outflows are detected in absorption both in the UV (through CIV and OVI) and X-rays (through hundreds of transitions from ionized metals, e.g. [14] and references therein), while in neither of our two cases we do detect associated OVI. Moreover photoionized X-ray outflows in absorption have not been seen in blazars (early reports of X-ray absorbers in blazars[15,16], have not been confirmed by later, higher spectral resolution, observations, e.g. [1,3,5]).

In the following we investigate this possibility further. If the absorber was part of the blazar environment, assuming a range of density of $10^6$-$10^{11}$ cm$^{-3}$ (typical of different gaseous components in an AGN environment, from the parsec-scale torus, to the sub-parsec scale Broad Emission Line Clouds -BELRs- or Warm Absorbers - WAs) and distances from 1 pc (the size of the torus for an object with the non-beamed luminosity of Mkn 421, i.e. L ~ $10^{42}$ erg s$^{-1}$) down to 0.5-1 light day (the size of the BELR and, possibly, the WA) the ionization parameter U (defined as the ratio between the density of ionizing photons and the gas density) at the illuminated face of the cloud, during 100 mCrab outbursts of the source, would reach values ranging between $U_{torus}$ = 30 and $U_{BELR-WA}$=0.3-30, putting most of the oxygen in the form of OVII and OVIII (for the BELR cloud) or OVIII and OIX (for the WA or the torus; see e.g. Fig. 1 in [17]). In our spectrum we only detect OVII, and measure upper limits for OVIII implying $U \lesssim 1$, so absorbing material from either the torus or a WA in the environment of Mkn 421 along our line of sight is unlikely to produce the observed features. A particularly dense BELR cloud, instead, could. However, typical equivalent H column densities for the BELR clouds are in the range $10^{22}$-$10^{24}$ cm$^{-2}$ (e.g. [18]), and metallicities are typically at least solar (e.g. [19]). This would imply OVII columns of $> 2 \times 10^{18}$ cm$^{-2}$ (assuming an OVII fraction of $f_{OVII}$ = 0.2 at $U_{BELR}$=0.3, see Fig. 1 in [17]), more than 3 orders of magnitude larger than the value we actually measure.

The same argument can be used to estimate the ionization degree of an extremely high velocity ISM cloud, located at kpc-scale from the blazar's host galaxy center (and along the line of sight to Mkn 421). Assuming typical ISM densities of $\lesssim 1$ cm$^{-3}$ we obtain U $\gtrsim$ 30 during outburst phases, which implies virtually no OVII, with $f_{OIX} = 0.95$ and $f_{OVIII} = 0.05$ (e.g. [17]). Finally, we also looked for variability of the strongest lines of these two systems in the two separate ACIS-LETG and HRC-LETG spectra of MKn 421, taken about 6 months apart. Variability would prove an AGN origin of the absorbers. We did not detect any variations in the equivalent width of the OVII lines of our systems (for details, please see Nicastro et al., 2004, in preparation). We conclude that the WHIM identification is the most likely for both the observed *z*=0.011 and *z*=0.027 systems.

## Metallicity Dependence on Modeling

Predictions of metal abundances in the WHIM are still highly uncertain, mostly due to the limited inclusion of galaxy-IGM or AGN-IGM feedback in current hydrodynamical simulations for the formation of structures. However values of about [O/H] = -1 or even slightly larger, are plausible at very low redshift[20], consistent with our estimates of [O/H] ~ -1 for both our WHIM systems. Our metallicity estimates (i.e. ratios of metals to H) depend only weakly on accurate modeling of the physical condition of the gas, since the exact position (i.e. temperature) and value of the, e.g., OVII/HI minimum in gas that is mainly collisionally ionized but still undergoing residual photoionization by the diffuse X-ray Background (XRB) and/or the blazar's radiation during powerful outbursts, itself depends only weakly on whether the gas is close to equilibrium (both with the ionizing photons and thermally). This dependence is instead much stronger for relative metallicity ratios (e.g. [N/O], [C/O] or [Ne/O]). In the case of our [N/O] estimate for the z=0.027 system, for example, the OVI/NVII and OVII/NVII ratio curves show a dramatic change in slope at logT ~ 6.4 (Nicastro et al., 2005, in preparation). The steepness of these curves at logT $\gtrsim$ 6.4 determines the [N/O] overabundance required, and depends critically on the baryon density. We adopt $n_b = 10^{-5}$ cm$^{-3}$, but at a density of $2 \times 10^{-6}$ cm$^{-3}$ a [N/O]=0 solution could be found for all ions except OVIII, at logT=6.2. The [N/O] overabundance also depends on whether the gas is in equilibrium or not, as recombination timescales are different for different elements and ions (e.g. [17]).

## References to Appendix B

1. Nicastro, F. et al., The far-ultraviolet signature of the `missing' baryons in the Local Group of galaxies, *Nature*, **421**, Issue 6924, 719-721, (2003)

2. Nicastro, F. et al., Chandra Discovery of a Tree in the X-Ray Forest toward PKS 2155-304: The Local Filament?, *Astrophys. J*, **573**, 157-167, (2002)

3. Fang, T., Sembach K.R., Canizares, C.R., Chandra Detection of Local O VII Heα Absorption along the Sight Line toward 3C 273, *Astrophys. J*, **586**, L49-L52, (2003)